\documentclass[sn-chicago, iicol]{sn-jnl}
\usepackage{lscape}
\usepackage{rotating}
\jyear{2021}
\begin{document}

\title[ ]{A wavelet-based detector function for characterizing intermittent velocity signals}

\author{\fnm{Satyajit} \sur{De}}\email{satyajitde@iisc.ac.in}

\author{\fnm{Aditya} \sur{Anand}}\email{adityaanand@iisc.ac.in}

\author{\fnm{Sourabh} \sur{S. Diwan}}\email{sdiwan@iisc.ac.in}

\affil{\orgdiv{Department of Aerospace Engineering}, \orgname{Indian Institute of Science}, \orgaddress{\city{Bengaluru}, \postcode{560012},  \country{India}}}

\abstract{In this work, we propose a new detector function based on wavelet transform to discriminate between turbulent and non-turbulent regions in an intermittent velocity signal. The derivative-based detector function, which is commonly used in intermittency calculation schemes, shows large fluctuations within turbulent parts of the signal and requires averaging over a certain ``smoothing period'' to remove the fake drop-outs, introducing subjectivity in calculating intermittency. The new detector function proposed here is obtained by averaging the ``pre-multiplied wavelet energy'' over the entire frequency range, resulting in a function that is much smoother than the derivative-based detector and at the same time has a good discriminatory property. This makes the choice of the smoothing period unnecessary and removes the subjectivity associated with it. We demonstrate the effectiveness of the wavelet detector within the framework of the widely-used method by Hedley and Keffer (1974, J. Fluid Mech., V64, pp625), for a range of velocity signals representing the different stages of roughness-induced transition. The wavelet detector function works well in detecting the edge intermittency of a canonical turbulent boundary layer as well, thereby highlighting its generality. Our detector can, in principle, be used with any method of specifying threshold for obtaining an indicator function.}


\maketitle

\section{Introduction}\label{sec1}

Intermittent velocity signals are observed in a variety of transitional and turbulent flows. In the context of wall-bounded flows, a prototypical example exhibiting large-scale intermittency is the transitional boundary layer which is characterized by presence of ``turbulent spots''. Turbulent spots are defined as islands of turbulence surrounded by quasi-laminar regions \citep{emmons_spot_1951} and are observed for both ``natural'' and ``bypass'' modes of transition \citep{Durbin2017}. From the onset of transition, turbulent spots grow in number and size, and eventually merge together to form a fully turbulent flow \citep{emmons_spot_1951}. Furthermore, intermittent velocity signals are also observed in the outer region of a turbulent boundary layer due to the effects of ``edge intermittency'', associated with its highly convoluted edge separating the turbulent and non-turbulent fluid \citep{Kovasznay1970}. Quantifying the intermittency of a velocity signal is important from both fundamental and applied perspectives; the former can help us understand how non-turbulent fluid acquires vorticity across a turbulent/non-turbulent interface and the latter can lead to better engineering models for transitional and turbulent flows.

For the transitional boundary layers, the intermittency is quantified by a parameter called the ``intermittency factor'' ($\gamma$) defined as the fraction of time for which the velocity signal is turbulent \citep{narashimha_intermittency_distribution1957}. Over the decades, several methods have been developed to detect the turbulent/non-turbulent interface and calculate the intermittency factor, many of which use one-dimensional velocity measurements from a hotwire probe. The turbulent part of the signal is visually discernible as a region of high frequency and high amplitude fluctuations. This aspect of the signal is enhanced by sensitising the signal usually by differentiating, double differentiating or high-pass filtering \citep{hedley_keffer_pdf_intermittency1974turbulent, fransson2005transition}; the sensitized signal is called the detector function. The detector function has a relatively higher magnitude in the turbulent part than in the non-turbulent part and an indicator function is obtained by applying a threshold on the magnitude of the detector function. The indicator function is assigned a value of 1 when the detector function is higher than the set threshold and 0 otherwise.  The time mean of the indicator function gives the intermittency factor, which varies from 0 to 1 over the extent of the transition zone \citep{narasimha_transition_1985}. When the detector function is obtained by differentiation, the turbulent part of the signal is accentuated showing large fluctuations, which results in some fake dropouts i.e., the indicator is 0 in the turbulent part of the signal. Similarly, the non-turbulent part can have spurious fluctuations where the indicator is assigned a value of 1. To rectify this issue, a smoothing process is applied on the detector function by averaging it over a fixed time interval, $T_s$, called the smoothing period \citep{hedley_keffer_pdf_intermittency1974turbulent}. The signal obtained after smoothing is called the criterion function and a threshold is now applied on the criterion function to obtain an indicator function. \cite{canepa2002experiences} and \cite{jahanmiri2011turbulent} have summarized some of the widely-used methods for calculating $\gamma$ proposed in the literature; a brief account of the important developments in this field can be found in a recent paper by \cite{veerasamy2020rational}.

The above procedure involves subjectivity at two levels -  first, the determination of the smoothing period and second, the determination of the threshold. A perusal of the literature shows that there is no unique specification of the smoothing period. Moreover, it is not clear what the appropriate time scale should be for scaling the smoothing period. \cite{hedley_keffer_pdf_intermittency1974turbulent} used a smoothing period of 4 times the sampling interval ($\Delta t$). This period was found to be of the order of the Taylor microscale and about 28 times the Kolmogorov time scale ($t_K$). Other authors have used a range for $T_s$ as $3-10 \Delta t$ in terms of the sampling period \citep{canepa2002experiences, veerasamy2020rational} or $15-200t_K$ in terms of the Kolmogorov scale \citep[see][and the refernecs therein]{canepa2002experiences}. \cite{anand2020time} found it necessary to use a much higher value of $T_s=100\Delta t$ in calculating $\gamma$ for their measurements on roughness-induced transition. Another way of expressing $T_s$ is in terms of the large-eddy time scale  $T_L=\delta/U_{\infty}$, where $\delta$ is the boundary layer thickness and $U_{\infty}$ is the freestream velocity. \cite{canepa2002experiences} have listed studies which have used a range of $T_s\approx1-6 T_L$; \cite{canepa2002experiences} themselves found values of $T_s = 3.7 T_L$ and $8 T_L$ to be more appropriate for the two Reynolds numbers used in their exercise on hotfilm signals, which were measured over gas-turbine blades. They found that the choice of $T_s$ had a significant effect on determining the intermittency distribution along the blade chord. \cite{samson2021instability} used $T_s = 2.5 T_L$ in calculating $\gamma$ for adverse-pressure-gradient boundary layers. The subjectivity in determining the detector/criterion function is also present for methods which use high-pass filtering of the velocity signal, as the filter cut-off frequency can be different for different types of signal \citep{fransson2005transition, Mamidala2022}. To the best of  our  knowledge, there does not exist any method that specifies the smoothing time or the filter frequency in an objective manner.

The second level of subjectivity has been addressed by several investigators by proposing various methods for determining the threshold on the detector/criterion function.  \cite{hedley_keffer_pdf_intermittency1974turbulent} proposed that the point of maximum curvature of the cumulative-distribution-function (CDF) of the criterion function can be chosen as the threshold for obtaining indicator function. The dual-slope method by \cite{kuan_wang_1990investigation} involves choosing the threshold as the point of intersection of two straight lines on the intermittency vs. threshold curve. \cite{nolan2013conditional}, determine the threshold by a method prescribed by \cite{otsu1979threshold} that maximizes the contrast in variance between the laminar and turbulent zones. More recently,  \cite{veerasamy2020rational} have proposed a method of selecting the maximum laminar perturbation near the onset of transition as the single threshold to be used for all the locations within the transition zone. This choice is based on the assumption that the laminar perturbations stay nearly constant throughout the transition zone and \cite{veerasamy2020rational} demonstrated the method for the transition induced by an aerofoil wake. It is not clear whether the method can be applied for transition induced by other types of disturbances. Some of the other methods of choosing a threshold are due to \cite{fransson2005transition} and \cite{Mamidala2022}. In most of these methods, the final determination of the threshold is based on a visual match between the indicator function and the signal (or criterion function) at some level; see \cite{jahanmiri2011turbulent}.

The present work deals with the first level of subjectivity. We propose a new detector function based on wavelet transform that does not require specification of smoothing period. Wavelet transform is a commonly used tool to study time-frequency behaviour of velocity signals and has been used before for transitional boundary layers especially in the context of conditional sampling of the turbulent and non-turbulent parts of the signal \citep{Volino2005, zhang2018conditional}. \cite{simoni2016wavelet} used wavelet transform to define an intermittency measure based on the count of ``events'' (representing vorticies) in two-dimensional snapshots of the velocity field; this method, however, is specific to 2-D flow fields and cannot be applied on velocity time series. We are not aware of any other study that has used wavelet-based method for determining a detector function. The present idea is based on the observation that the non-turbulent portions in an intermittent velocity signal are associated with extended regions in the time-frequency plane where the magnitudes of wavelet coefficients are low \citep{anand2020time}. The detector function is then obtained by simply averaging  the ``pre-multipied'' wavelet energy over the entire frequency axis, obviating the need for any additional smoothing. We demonstrate the utility of this approach on transitional velocity signals measured in our previous work on roughness-induced transition  \citep{anand2020time}. Furthermore, the new detector function works well for detecting edge intermittency for velocity signals measured in a fully turbulent (canonical) boundary layer, suggesting its general applicability.

The paper is organized as follows. Section 2 briefly describes the experimental setup used for measuring velocity in a transitional boundary layer. The subjectivity associated with the specification of a smoothing period is discussed in section 3. The new wavelet-based detector function is proposed in section 4 and is demonstrated for transitional velocity signals therein. In section 5 we test the new detector function on intermittent velocity signals in canonical wall turbulence. Conclusions are presented in section 6.

\section{Experimental setup for roughness-induced transition}\label{sec2}

The velocity signals used for the present analysis were measured in our previous experimental work on roughness-induced transition \citep{anand2020time}. Here we provide a brief outline of the experimental setup for the sake of completeness. The experiments were conducted in a low-speed wind tunnel at Indian Institute of Science, Bengaluru. The wind tunnel has a test-section of 0.5 m x 0.5 m (and 3 m length) and has a free-stream turbulence intensity of about 0.1\%  \citep{anand2020time}.  

 \begin{figure*}[h]%
    \centering
    \includegraphics[width=\textwidth]{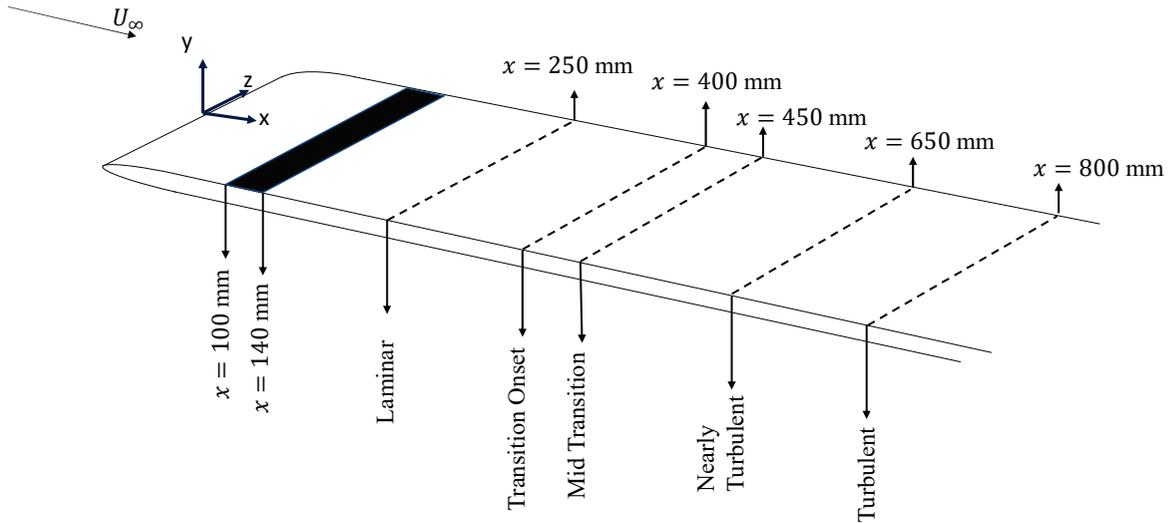}
    \caption{Schematic of the flat plate used for generating a transitional boundary layer downstream of a 40 mm wide, grade 24 (coarse) roughness strip, shown in solid hatching. The co-ordinate system used herein is also shown.}\label{Plate}
\end{figure*}

The experiments were performed on a flat plate, 6.25 mm thick and 2.1 m long, and having a super elliptic leading edge. A 40 mm wide spanwise strip of grade 24 emery paper was used as a randomly distributed surface roughness (overall roughness height=1.5 mm) to induce transition and it was placed 100 mm downstream of the leading edge as shown in Fig. \ref{Plate}. Dantec dynamics 55P11 miniature boundary-layer hotwire probe was used to measure instantaneous velocity signals at a sampling rate of 20 kHz. The sampled data was low-pass filtered, using a hardware filter, at a cut-off filter of 10 kHz to remove aliasing errors as per the Nyquist criterion. The output voltage from the hotwire probe was corrected for ambient temperature variations \citep{bruun1996hot} and calibrated against Pitot-tube velocities using the King’s law.  More details on the experimental setup and measurement chain can be found in \cite{anand2020time}.

\begin{figure*}[h]%
\centering
\includegraphics[width=\textwidth]{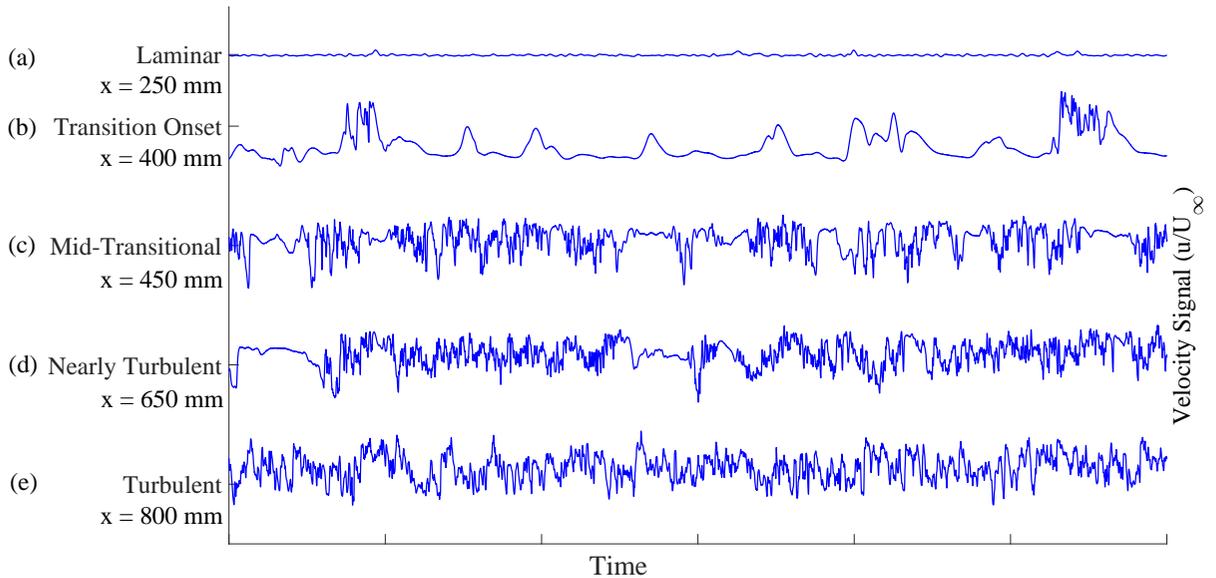}
\caption{Velocity signals from various stages of transition along with their corresponding $x$-locations; see Fig. \ref{Plate}.}\label{Signals}
\end{figure*}

\cite{anand2020time} reported velocity profiles measured over the entire transition zone downstream of the roughness at $U_{\infty}=7.5$ m/s with a roughness Reynolds number of $U_{\infty}k/\nu=640$, where $k$ is the maximum roughness height and $\nu$ is kinematic viscosity. For the present analysis we have chosen five velocity signals representative of the various stages of transition. The streamwise locations for these signals are shown schematically in Fig. \ref{Plate} and are listed in Table \ref{tab1}; the velocity signals are shown in  Fig. \ref{Signals} (a)-(e). The signal duration used for our analysis is 30 s but we show a short stretch of time (0.6 s) in Fig. \ref{Signals} to bring out the contrast between the turbulent and non-turbulent parts. The wall-normal locations of the signals are in the range $y/\delta \approx 0.1-0.4$, which is the typical region over which the intermittency factor is approximately constant \citep{Matsubara1998}. \cite{veerasamy2020rational} sampled velocity signals at $y/\delta^*=0.5$ for their analysis, which falls approximately in the above range over the extent of the transition zone. Table \ref{tab1} also lists values of $\gamma$ calculated based on the Hedley-Keffer method \citep{hedley_keffer_pdf_intermittency1974turbulent} described in the next section. Based on the values of $\gamma$, we denote the signals as ``Laminar'', ``Transition Onset'', ``Mid-Transitional'', ``Nearly Turbulent'' and ``Turbulent'' (Table \ref{tab1}).

\begin{table*}[h!]
\begin{center}
\begin{minipage}{290pt}
\caption{Streamwise locations, intermittency values and stages of transition for the velocity signals chosen for the present analysis.} \label{tab1}%
\begin{tabular}{@{}ccccc@{}}
\toprule
Case & $x$ location (mm) &  Intermittency ($\gamma$) & Transition Stage\\
\midrule
a &	250 &	$\approx 0$ &	Laminar \\
b &	400 &	0.09 &	Transition Onset \\
c &	450 &	0.42 &	Mid-Transitional \\
d &	650 &	0.87 &	Nearly Turbulent \\
e &	800 &	1 &	Turbulent \\
\botrule
\end{tabular}
\end{minipage}
\end{center}
\end{table*}

\begin{figure*}[h]%
\centering
\includegraphics[width=\textwidth]{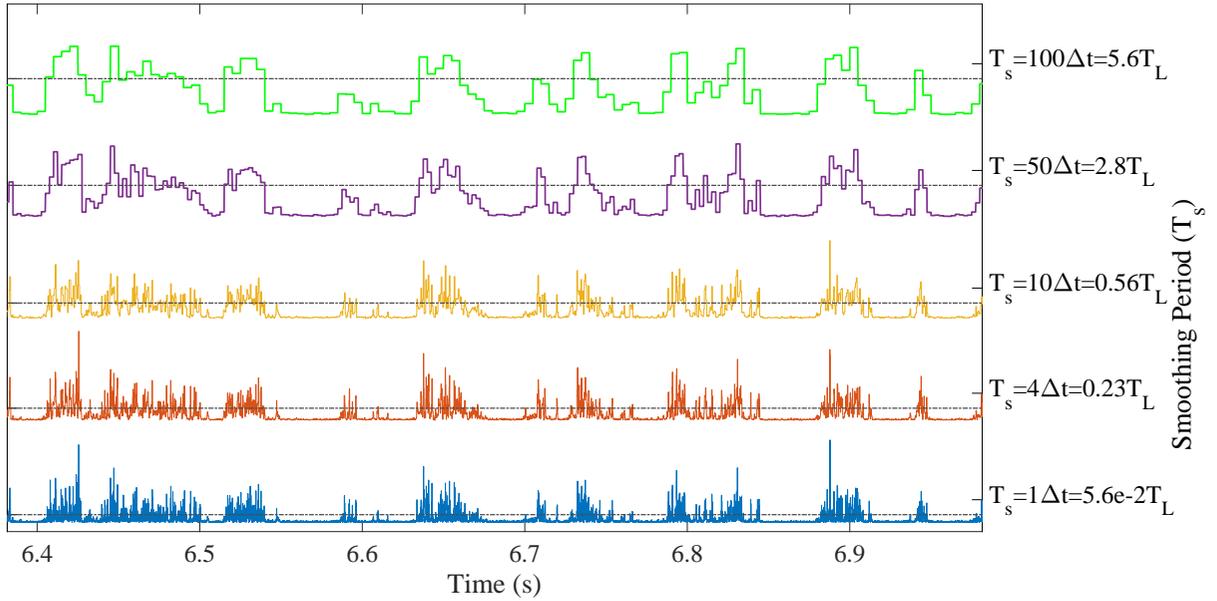}
\caption{Effect of smoothing period on the criterion function for the Mid-Transitional case (Table \ref{tab1}). The detector function is averaged over the time interval $T_s$. The dash-dotted lines show the threshold as calculated by the Hedley-Keffer method (i.e. the maximum curvature of the CDF) for each criterion function $\Delta t$ is the sampling interval and $T_L$ is the large-eddy time scale.}\label{Criterion}
\end{figure*}

\begin{figure*}[h]%
\centering
\includegraphics[width=\textwidth]{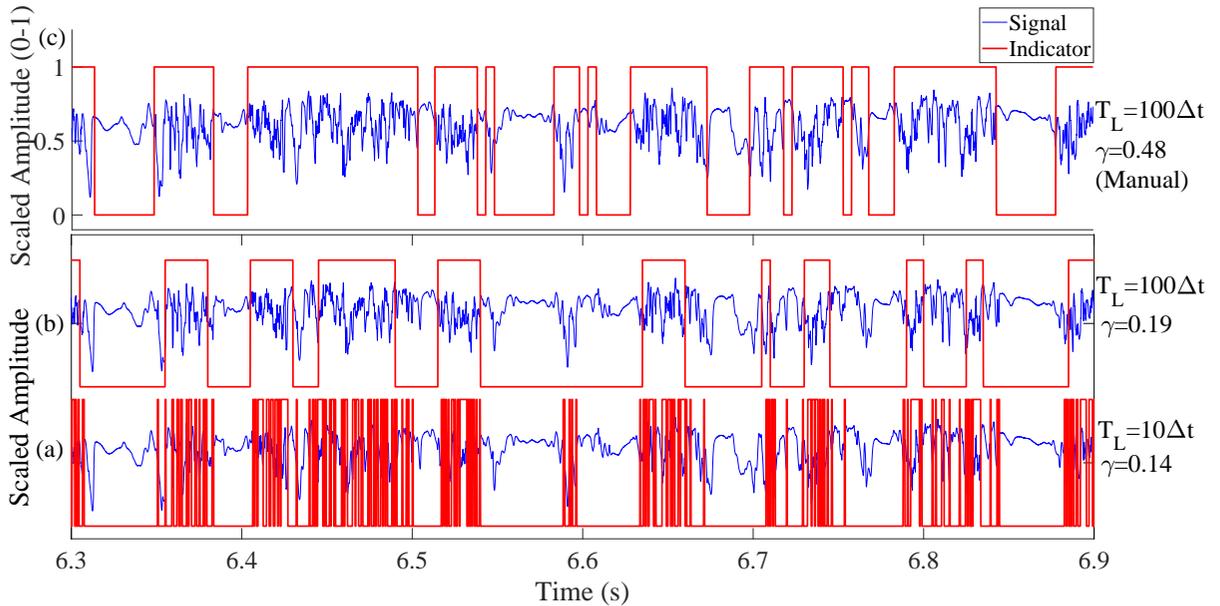}
\caption{(a,b) Indicator functions obtained for two different smoothing periods ($T_s=10\Delta t$ and $T_s=100\Delta t$) by using the Hedley-Keffer method \citep{hedley_keffer_pdf_intermittency1974turbulent}. (c) Indicator function determined using a manual threshold based on visual inspection.}\label{Indicator}
\end{figure*}

\begin{figure*}[h]%
\centering
\includegraphics[width=0.9\textwidth]{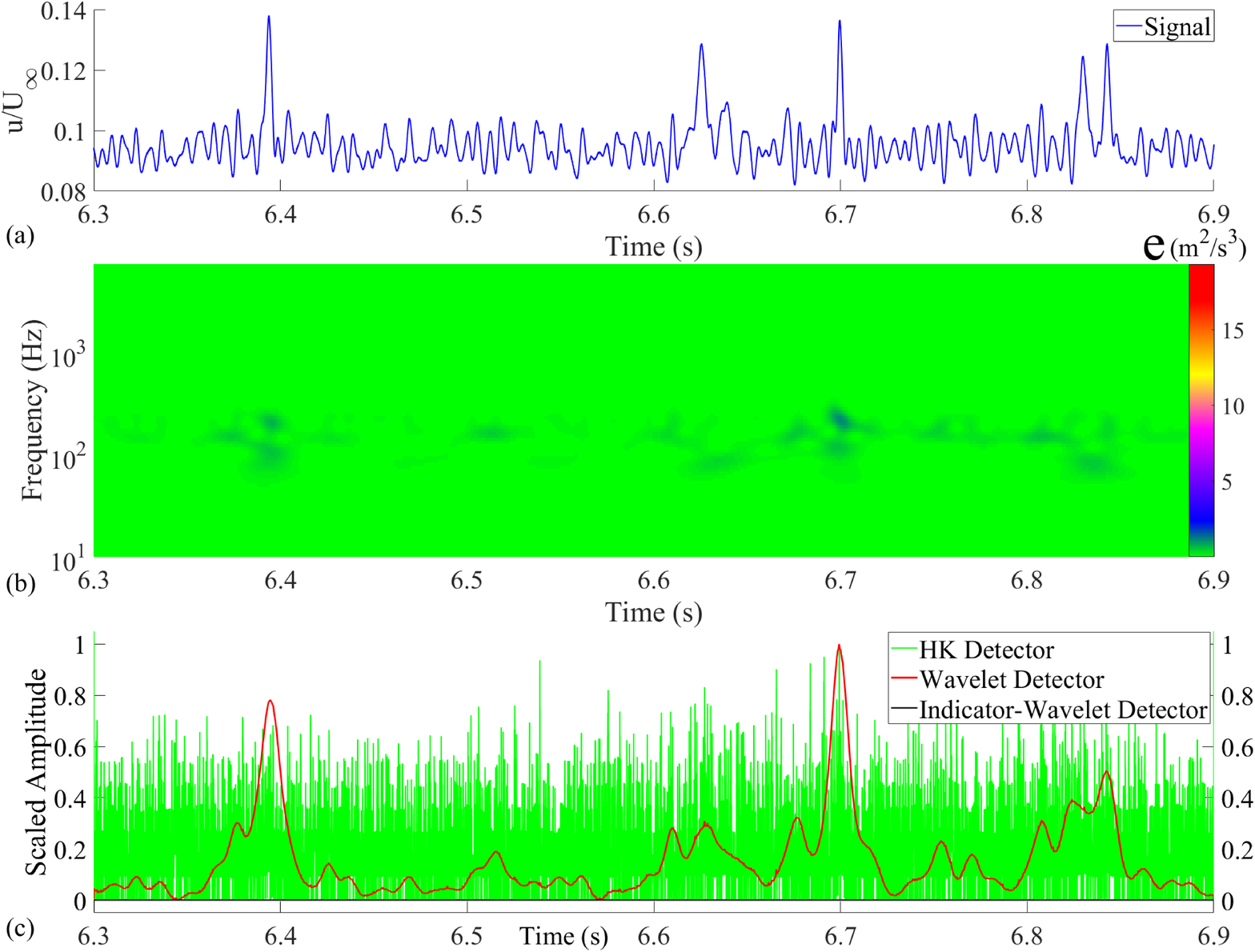}
\caption{Laminar case: (a) Velocity signal $(u/U_\infty)$ vs. time (s). (b) Contour plot of (pre-multiplied) wavelet energy ($fC_w^2)$ in frequency-time plane. (c) Comparison between the wavelet detector \textcolor{red}{(red line)} and the Hedley-Keffer (HK) detector \textcolor{green}{(green line)}. Both the detectors are scaled from 0-1 over the time interval shown. The indicator function obtained from the wavelet detector (``Indicator-Wavelet Detector'') is also shown \textcolor{black}{(black line).}}\label{Laminar}
\end{figure*}

\section{Effect of smoothing period on the calculation of intermittency factor}\label{sec3}

We first calculate intermittency factors for the signals in Fig. \ref{Signals} using the Hedley-Keffer method, and highlight the subjectivity introduced by the smoothing time on the determination of the criterion function. We choose the Mid-Transitional signal for outlining the step-by-step procedure. The absolute value of the double derivative of the velocity signal is chosen as the detector function, consistent with \cite{hedley_keffer_pdf_intermittency1974turbulent} and \cite{veerasamy2020rational}. Figure \ref{Criterion} plots the criterion functions obtained by using different smoothing periods:  $1\Delta t$ to $100\Delta t$. As can be seen, the criterion function gets progressively smoother with the increase in $T_s$. The smoothing period is also expressed in terms of the large-eddy time, $T_L$ (Fig. \ref{Criterion}). For the present exercise, we find $T_s=100\Delta t$ to be the appropriate smoothing period. This translates into $T_s=5.6T_L$ which falls in the range $T_s = 3.7-8T_L$ used by \cite{canepa2002experiences} for their analysis. 

To determine the threshold on the criterion function, we use the CDF method proposed by \cite{hedley_keffer_pdf_intermittency1974turbulent}. The maximum curvature of the CFD of the criterion function is used as a first approximation for the threshold. This threshold is a function of the smoothing period and is plotted in Fig. \ref{Criterion} as dash-dot lines. The indicator functions obtained from these threshold values are plotted in Fig. \ref{Indicator} for two smoothing periods: $T_s=10\Delta t$ and $100 \Delta t$, superposed on the velocity signals. It is clear from Fig. \ref{Indicator}(a) that there are several fake drop-outs within the turbulent regions at $T_s=10\Delta t$, as a result of which the value of $\gamma$ is quite low. At $T_s=100\Delta t$ there is an improved match between the indicator and the signal, although fake drop-outs are still present (Fig. \ref{Indicator}(b)). Finally, the threshold was manually adjusted to get a good match between the indicator function and the signal by visual inspection \citep{jahanmiri2011turbulent}, which is shown in Fig. \ref{Indicator}(c). This ensures that there are no fake 0s or spurious 1s in the indicator function. The resulting value of $\gamma (=0.48)$ is much higher than those obtained for the CDF-based threshold (Fig. \ref{Indicator}). (Note that in this exercise and in what follows, $\gamma$ is calculated by averaging the indicator function over the entire stretch of the signal, i.e., 30 s.) The strong dependence of intermittency factor on the smoothing period seen here is consistent with the observation in \cite{canepa2002experiences} for flow past a turbine blade. It is not clear whether the sensitive dependence of $\gamma$ on $T_s$ is present only for certain types of transitional flows and using specific methods; see, for example, \cite{veerasamy2020rational} and \cite{samson2021instability}. The values of $\gamma$ listed in Table \ref{tab1} for other velocity signals have been obtained by using the same procedure, i.e. with $T_s = 100 \Delta t$ and threshold determined by visual inspection. The above exercise brings out the subjectivity introduced by the smoothing period and threshold level in the accurate determination of the intermittency factor. 

In the next section, we propose a new wavelet-based detector function that largely eliminates the uncertainty associated with the specification of the smoothing period. We use the Hedley-Keffer method as the framework for demonstrating effectiveness of the new detector function.

\section{New detector function based on wavelet transform}\label{subsec2}

\begin{figure*}[h]%
\centering
\includegraphics[width=0.9\textwidth]{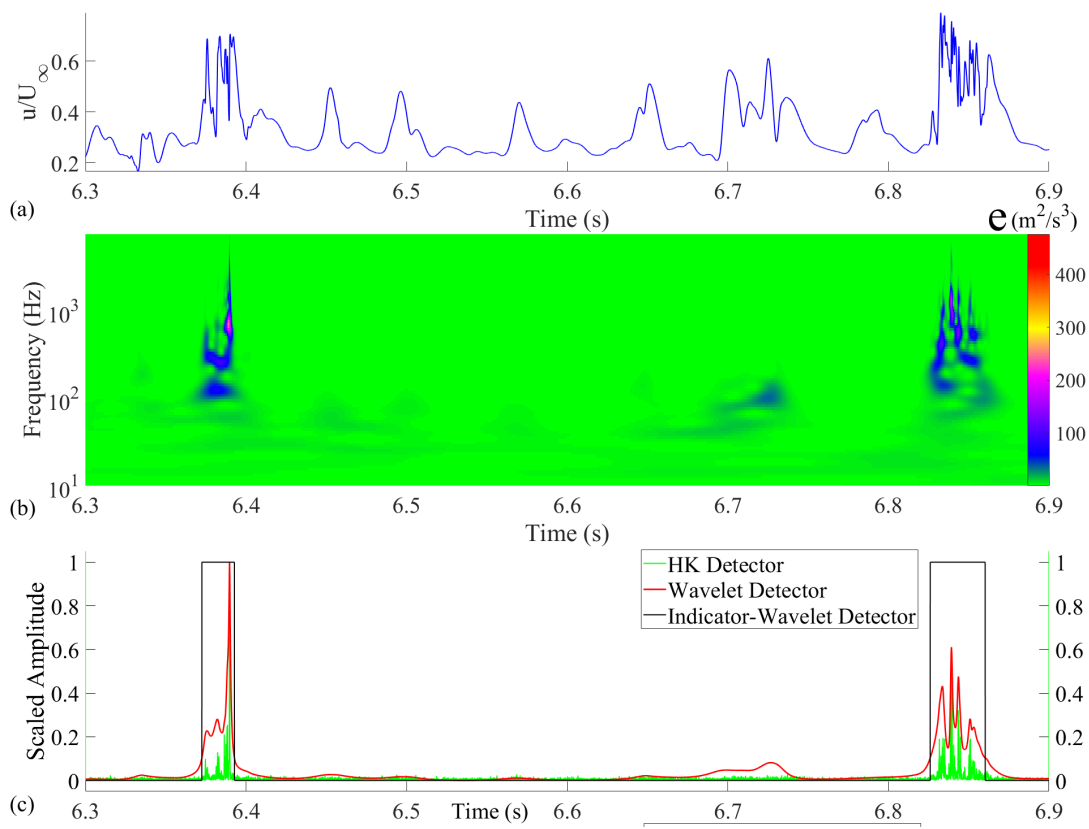}
\caption{Transition Onset case: (a) Velocity signal $(u/U_\infty)$ vs. time (s). (b) Contour plot of (pre-multiplied) wavelet energy ($fC_w^2)$ in frequency-time plane. (c) Comparison between the wavelet detector \textcolor{red}{(red line)} and the Hedley-Keffer (HK) detector \textcolor{green}{(green line)}. Both the detectors are scaled from 0-1 over the time interval shown. The indicator function obtained from the wavelet detector (``Indicator-Wavelet Detector'') is also shown \textcolor{black}{(black line).}}\label{Transition_Onset}
\end{figure*}

\begin{figure*}[h]%
\centering
\includegraphics[width=0.9\textwidth]{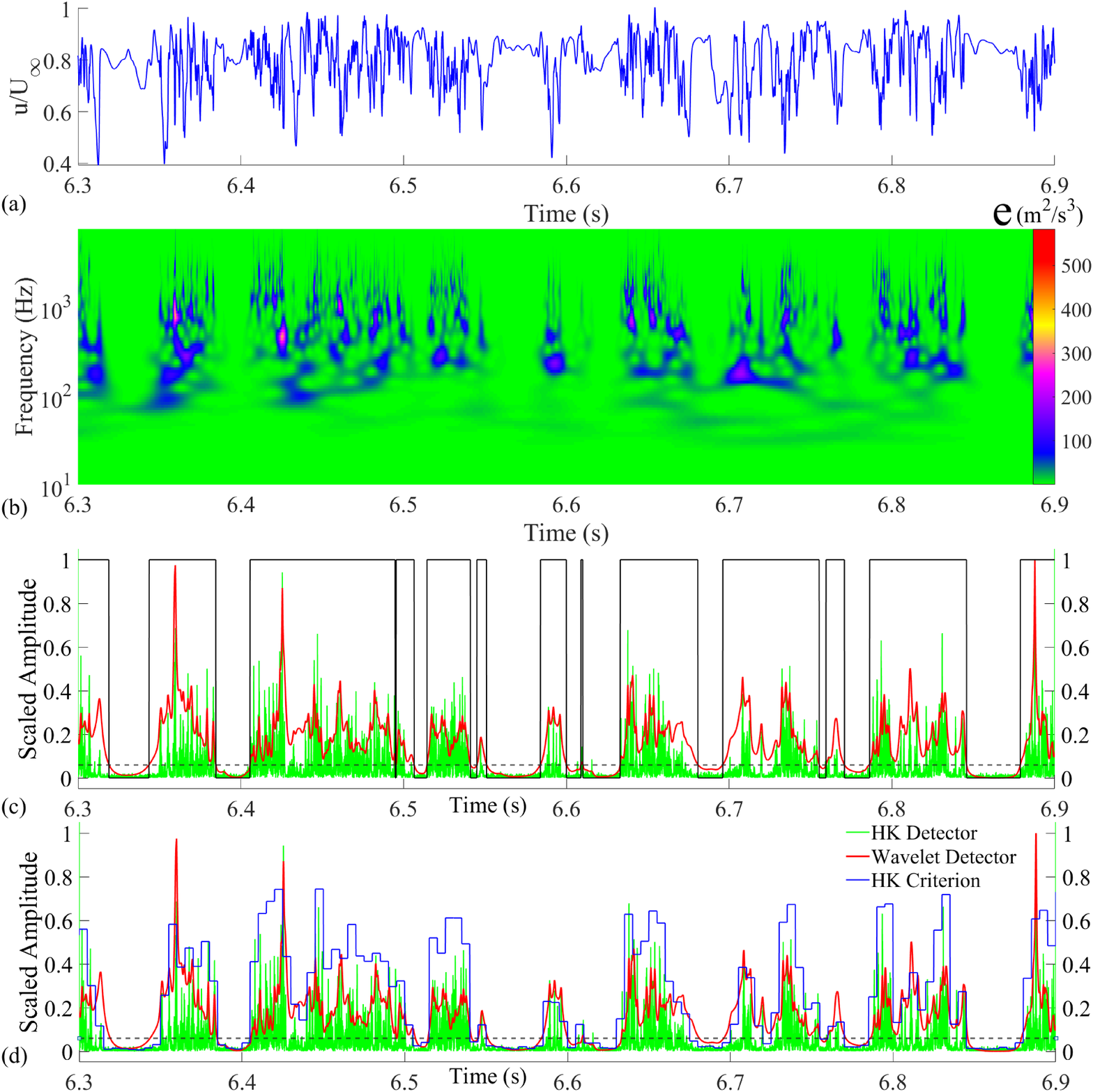}
\caption{Mid-Transitional case: (a) Velocity signal $(u/U_\infty)$ vs. time (s). (b) Contour plot of (pre-multiplied) wavelet energy ($fC_w^2)$ in frequency-time plane. (c) Comparison between the wavelet detector \textcolor{red}{(red line)} and the Hedley-Keffer (HK) detector \textcolor{green}{(green line)}. Both the detectors are scaled from 0-1 over the time interval shown. The indicator function obtained from the wavelet detector (``Indicator-Wavelet Detector'') is also shown \textcolor{black}{(black line).} (d) Comparison between the wavelet detector function \textcolor{red}{(red line)} and the HK criterion function \textcolor{blue}{(blue line)} for the smoothing period of $100\Delta t$; see Fig.\ref{Criterion}.}\label{Mid-Transitional}
\end{figure*}

Wavelet transform maps a signal in time, onto a time-frequency plane using functions called ``mother wavelet'' and their scaled and translated versions, as a basis for the mapping \citep{farge1992wavelet}. Since wavelet functions are band-limited both in time and frequency, the wavelet transform is ideal for time-frequency analysis of an intermittent signal, and its ability to detect turbulent spots in a transitional signal has already been demonstrated \citep{anand2020time, samson2021instability, zhang2018conditional};  see \cite{anand2020time} and \cite{anand2020thesis} for more details. In this work we have used the MATLAB command ``cwt'' to calculate the wavelet coefficients ($C_w$) for our analysis; ``cwt'' uses convolution algorithm to compute the continuous wavelet transform coefficients. The analytical Morlet wavelet is used as the mother wavelet \citep{anand2020time, zhang2018conditional}.

\begin{figure*}[h]%
\centering
\includegraphics[width=0.9\textwidth]{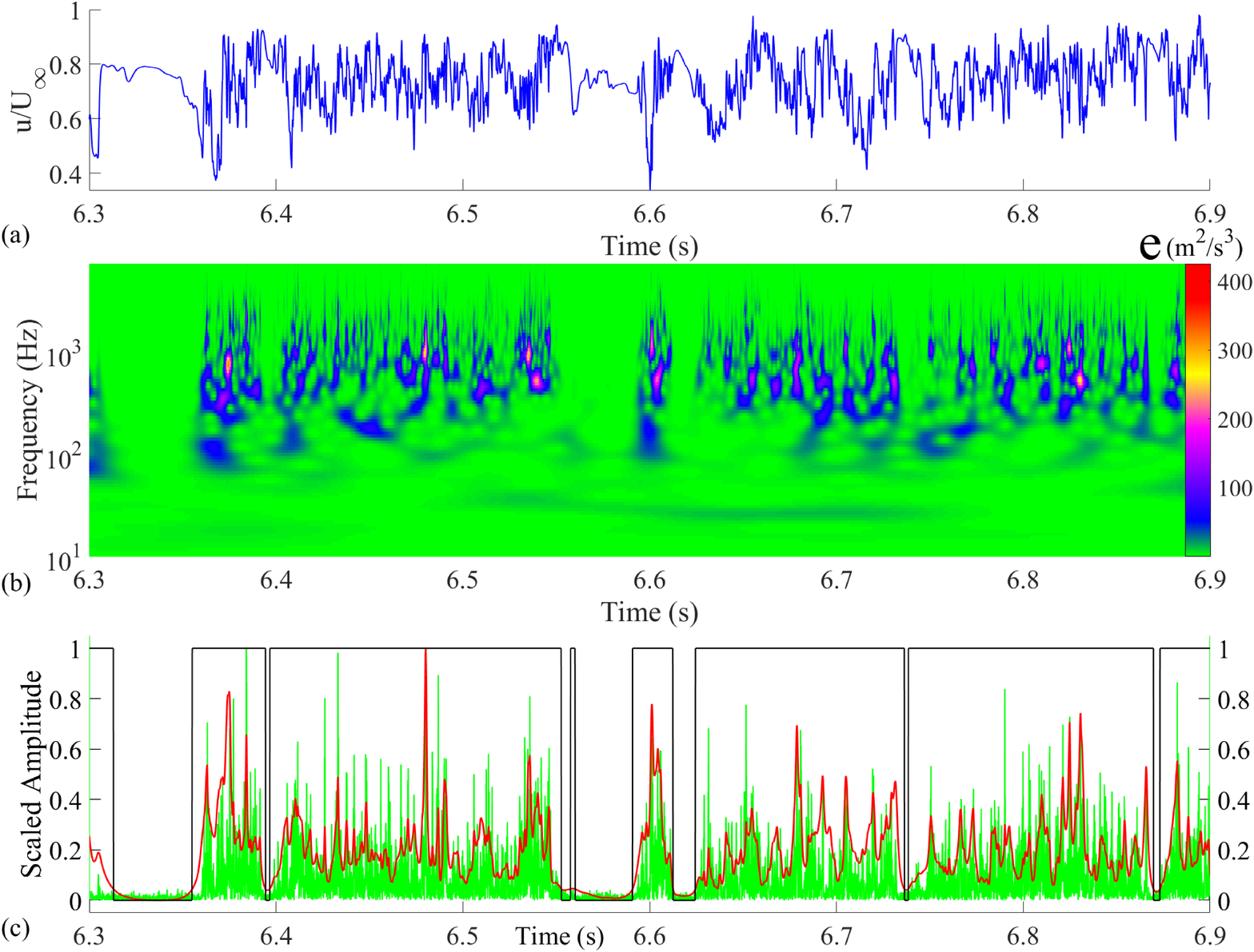}
\caption{Nearly Turbulent case: (a) Velocity signal $(u/U_\infty)$ vs. time (s). (b) Contour plot of (pre-multiplied) wavelet energy ($fC_w^2)$ in frequency-time plane. (c) Comparison between the wavelet detector \textcolor{red}{(red line)} and the Hedley-Keffer (HK) detector \textcolor{green}{(green line)}. Both the detectors are scaled from 0-1 over the time interval shown. The indicator function obtained from the wavelet detector (``Indicator-Wavelet Detector'') is also shown \textcolor{black}{(black line).}}\label{Nearly_Turbulent}
\end{figure*}

\begin{figure*}[h]%
\centering
\includegraphics[width=0.9\textwidth]{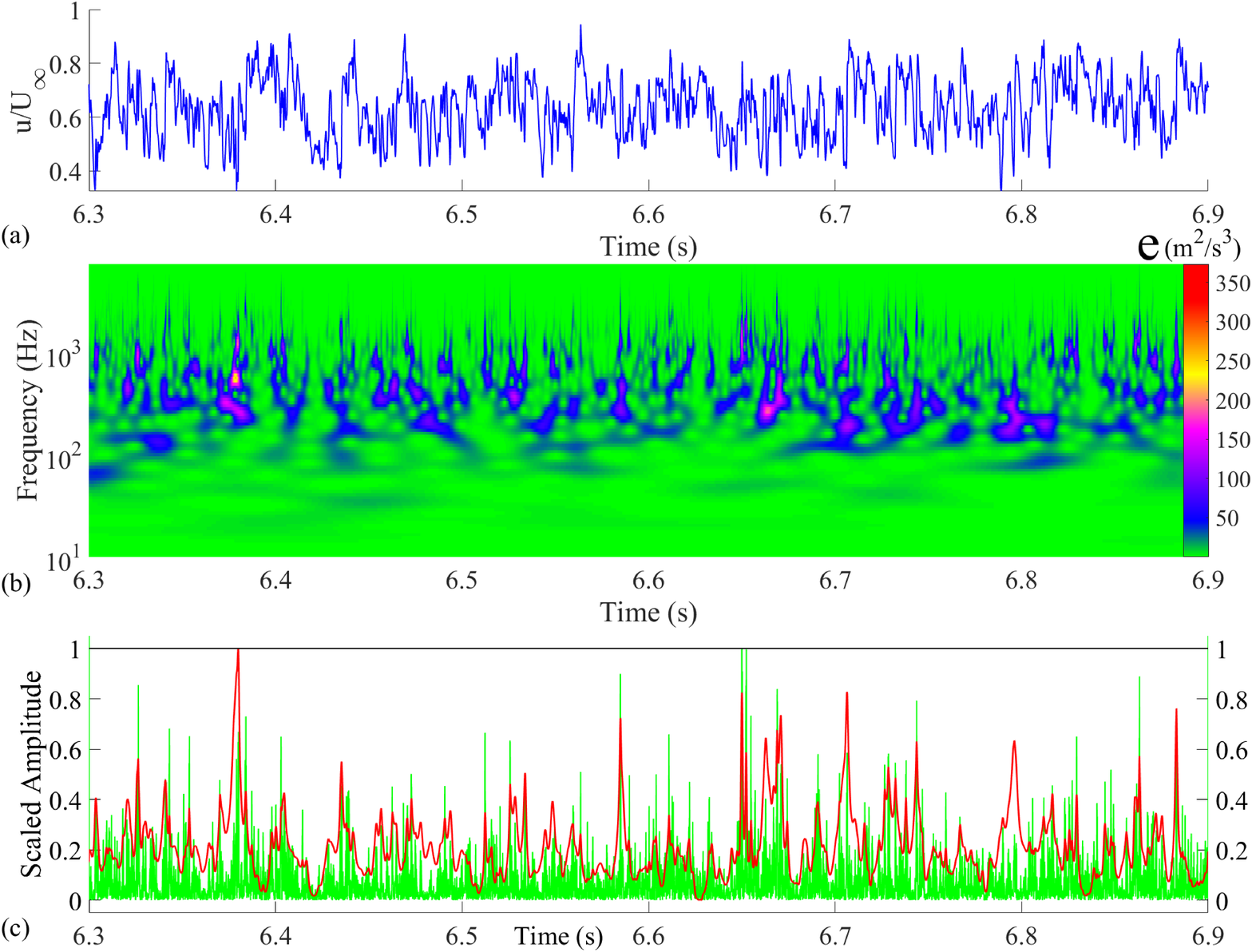}
\caption{Turbulent case: (a) Velocity signal $(u/U_\infty)$ vs. time (s). (b) Contour plot of (pre-multiplied) wavelet energy ($fC_w^2)$ in frequency-time plane. (c) Comparison between the wavelet detector \textcolor{red}{(red line)} and the Hedley-Keffer (HK) detector \textcolor{green}{(green line)}. Both the detectors are scaled from 0-1 over the time interval shown. The indicator function obtained from the wavelet detector (``Indicator-Wavelet Detector'') is also shown \textcolor{black}{(black line).}}\label{Turbulent}
\end{figure*}

The results from wavelet analysis for the five cases listed in Table \ref{tab1} are plotted in Fig. \ref{Laminar}-\ref{Turbulent}. In each figure, window (a) plots the non-dimensional velocity signal $(u/U_\infty)$ as a function of time, window (b) plots the corresponding squared wavelet coefficients multiplied with frequency ($fC_w^2=e$) \citep{anand2020time}, hereafter referred to as wavelet energy, as a function of time and frequency and window (c) plots wavelet-based detector function (described in the following) compared with the Hedley-Keffer detector function. Also plotted in window (c) is the indicator function obtained from the wavelet detector. The use of the pre-multiplied form of wavelet energy enables an even representation of wavelet energy on the logarithmic frequency axis, akin to the pre-multiplied Fourier spectrum commonly used in the wall turbulence literature \citep{McKeon2011}. The time window of 6.3 s to 6.9 s is chosen as it gives a better visual representation of these signals.

\begin{figure*}[h]%
\centering
\includegraphics[width=0.9\textwidth]{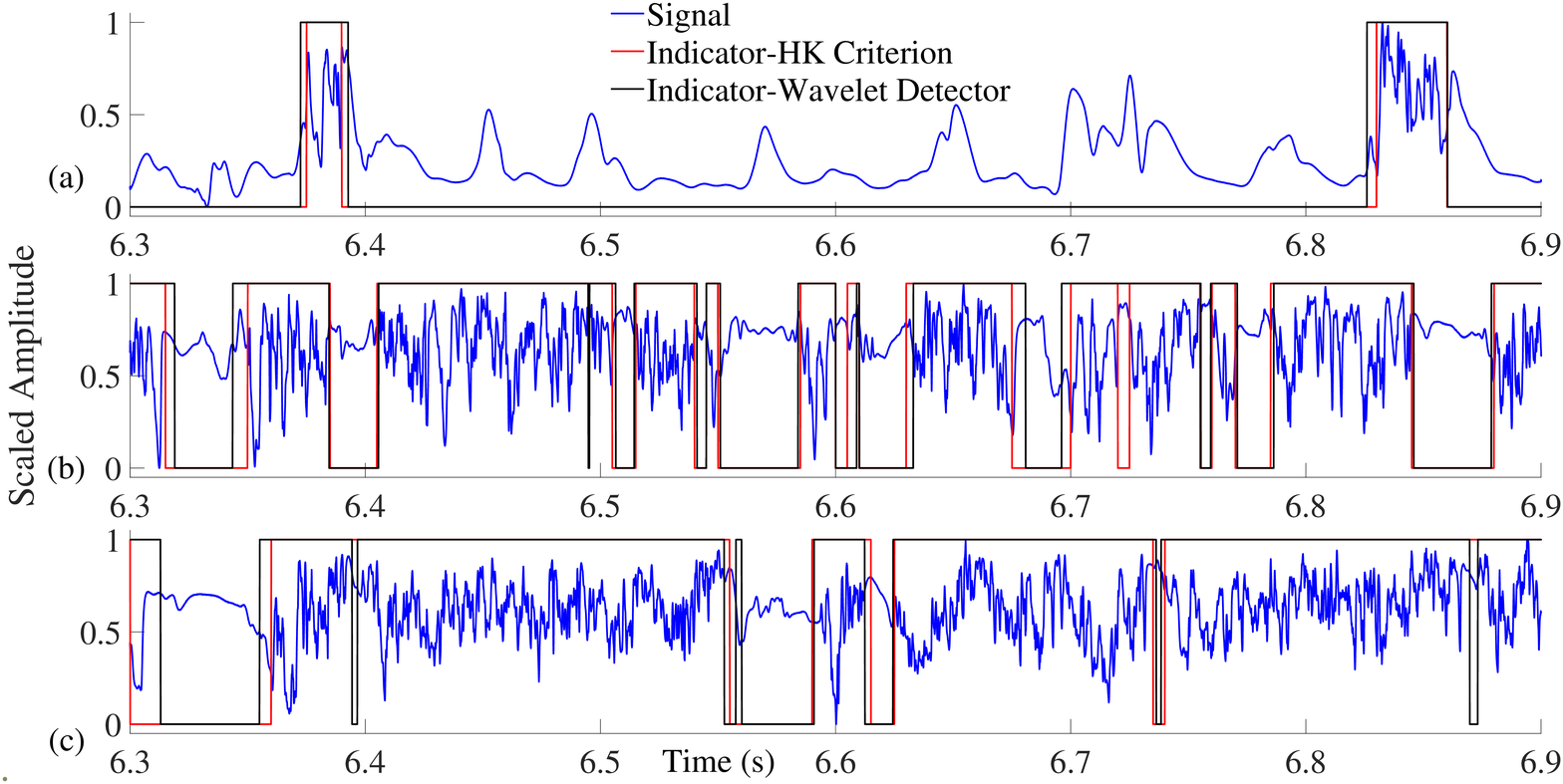}
\caption{Indicator functions obtained from the wavelet detector and HK criterion functions, superposed on the velocity signal. This shows the ``fit'' of the indicator on the signal. Cases: (a) Transition Onset (b) Mid-Transitional (c) Nearly Turbulent.}\label{Compare_Indicator}
\end{figure*}

To motivate the choice of the new detector function, we first discuss the Mid-Transitional signal (Fig. \ref{Mid-Transitional}) which was used for the detailed analysis of the Hedley-Keffer method in the previous section. It can be seen from Fig. \ref{Mid-Transitional}(b) that the presence of turbulent spots is associated with elevated levels of wavelet energy, localized in time, for frequencies higher than 100 Hz, whereas the non-turbulent regions are associated low wavelet energy as expected. The important observation from Fig. \ref{Mid-Transitional}(b) is that the non-turbulent parts of the velocity signal are represented as extended regions of low activity in the wavelet space  over the entire ``high'' frequency band, i.e., above $\approx 100$ Hz; this was also reported by \cite{anand2020time}. The energy contained in the ``low'' frequency band, i.e., $< 100$ Hz is more spread out along the time axis indicating that low-frequency motions are not as time-localized. Moreover, the wavelet energy associated with them is much smaller than that contained in the turbulent spots. This suggests that averaging the wavelet energy (in the pre-multiplied form) over the entire frequency band (Eq. \ref{detector}) should result in a time signal that brings out the contrast between the turbulent and non-turbulent regions fairly well and therefore serves as a discriminator between the two. The averaged wavelet energy is given as

\begin{equation} \label{detector}
D(t)=\frac{1}{n}\ \sum^{n}_{i=1}  e(t,f_i)
\end{equation}
where, $f_i$ is the set of discrete frequencies and $n$ is the total number of frequencies in the wavelet transform. We propose to use $D(t)$ in Eq. \ref{detector} as the new detector function, which is plotted in Fig. \ref{Mid-Transitional}(c) labelled as ``Wavelet Detector''. A comparison of the wavelet detector with the Hedley-Keffer detector (labelled as ``HK Detector''; Fig. \ref{Mid-Transitional}) shows that the former resembles the latter in overall behaviour but is much smoother and exhibits far less fluctuations than the latter. (Note that the detector functions shown in Fig. \ref{Mid-Transitional}(c) are scaled by their maximum values within the time interval 6.3-6.9 s.) In fact the wavelet detector seems to provide an approximate envelope for the HK detector in a way that a clear distinction between turbulent and non-turbulent parts can be made. This is also seen in Fig. \ref{Mid-Transitional}(d) where we compare the wavelet detector function and the Hedley-Keffer criterion function (``HK Criterion'') obtained for $T_s=100\Delta t$ (Fig. \ref{Criterion}); a fairly good resemblance is seen between the two functions. More importantly, the wavelet detector and HK criterion functions match well in the region close to the threshold level set for determining the indicator function, shown in Figs. \ref{Mid-Transitional}(c) and (d) by a dashed line. This threshold is the same as that obtained in the previous section using the Hedley-Keffer method, followed by a visual comparison between the HK criterion and the velocity signal (Fig. \ref{Criterion}c). Since the HK criterion and wavelet detector functions cross the threshold at nearly the same points (Fig. \ref{Mid-Transitional}c), one can expect that the indicator functions obtained from these two functions should be nearly identical; see Fig. \ref{Compare_Indicator}(b). The indicator function determined by using the wavelet detector (labelled as ``Indicator-Wavelet Detector'') is shown in Fig. \ref{Mid-Transitional}(c) by a black solid line.

\begin{figure*}[!h]%
\centering
\includegraphics[width=0.7\textwidth]{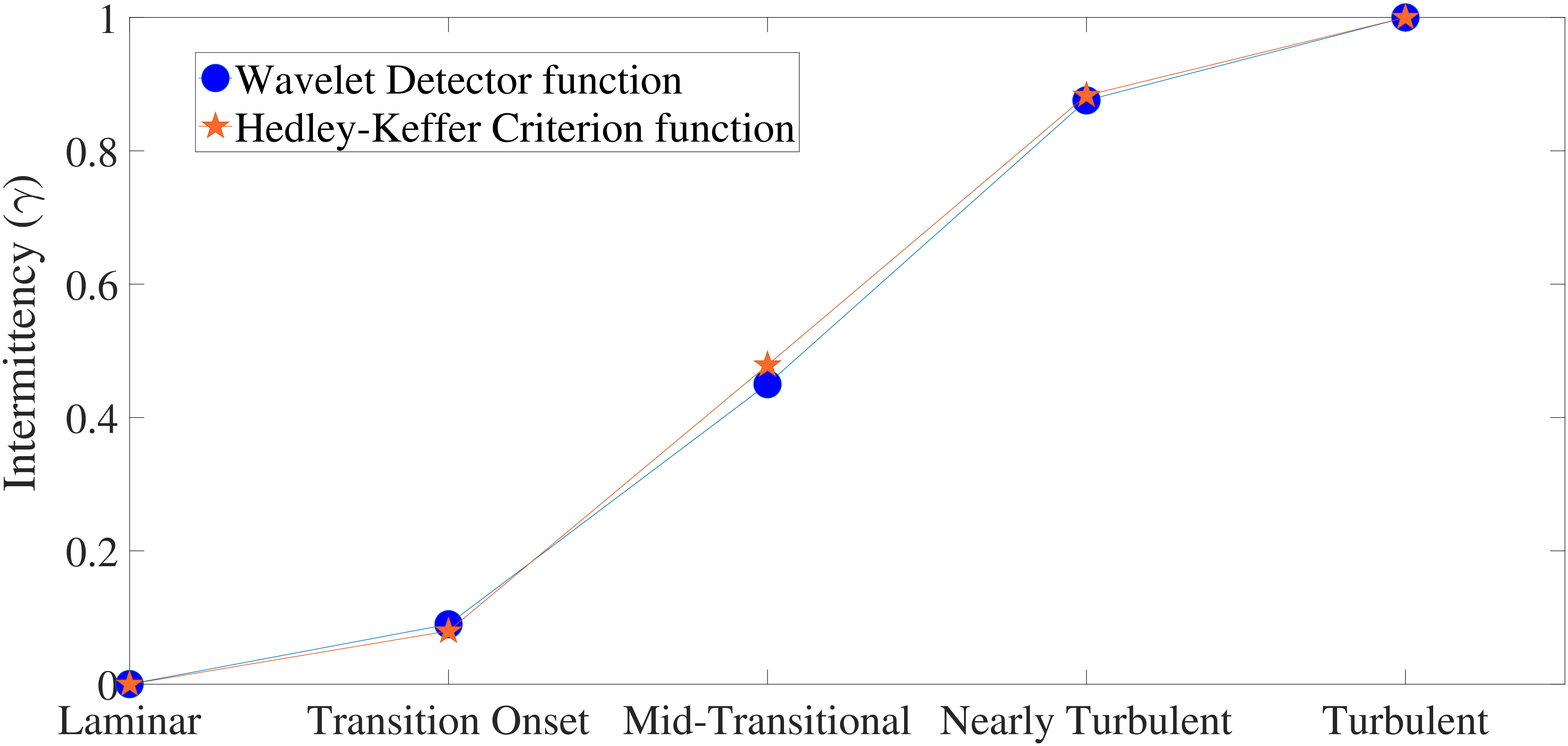}
\caption{Comparison between intermittency factors calculated based on the Hedley-Keffer criterion function and the wavelet detector function. A common threshold is employed for both the methods for a given stage of transition.}\label{Indi}
\end{figure*}

Next we discuss how the wavelet detector fairs for the velocity signals at other stages of transition (Fig. \ref{Signals}). For the laminar signal, the wavelet energy is low and is predominantly present around 100 Hz (Fig. \ref{Laminar}(b)). The HK detector is noisy whereas the wavelet detector is much smoother (Fig. \ref{Laminar}(c)). For this case, however, the latter does not act as an envelope over the former. This is because it is difficult to calculate intermittency for near-laminar signals accurately, which is a well-recognized fact in the literature \citep{jahanmiri2011turbulent}. The indicator function for this signal is zero at all times (Fig. \ref{Laminar}(c)). For the Transition Onset case we see emergence of localized wavelet energy for frequencies higher than 100 Hz (Fig. \ref{Transition_Onset}) and a clear distinction between the turbulent and non-turbulent parts; see Fig. \ref{Transition_Onset}(c). As we move downstream in the transition zone, the duration of turbulent spots increases which is also reflected in the wavelet-energy distribution; however the extended high-frequency regions associated with non-turbulent parts in the velocity signals are clearly seen at all stages of transition (Figs. \ref{Transition_Onset}(b) - \ref{Nearly_Turbulent}(b)). Consequently, the wavelet detector acts as a smoother version of the HK-detector at every stage of transition (after its onset) and is a better discriminator between the turbulent spots and non-turbulent regions (Figs. \ref{Transition_Onset}(c) and \ref{Nearly_Turbulent}(c)). This is also seen for the fully turbulent case for which non-turbulent parts are absent; the wavelet detector still picks up the variation of the HK-detector quite well (Fig. \ref{Turbulent}). The indicator functions plotted in Figs. \ref{Laminar}(c)-\ref{Turbulent}(c) are obtained by using the same threshold as used in the Hedley-Keffer method for each case(section 3), which is applied after scaling the wavelet detector from 0 to 1. It is important  that the indicator function shows a good fit on the detector function as the latter is the actual discriminator between the laminar and turbulent zones; Figs. \ref{Transition_Onset}(c)-\ref{Turbulent}(c) show that this is indeed the case for the wavelet detector.

The above exercise shows that the wavelet detector function provides an effective substitute for the Hedley-Keffer criterion function. This eliminates the need to introduce an additional step of averaging the detector function over an arbitrary value of the smoothing period that can only be determined empirically. Although there is averaging involved for the wavelet detector function also, it is performed over the frequency axis (instead of time axis) and the averaging interval is the \textit{entire} frequency range thereby eliminating any arbitrariness.

\begin{figure*}[h]%
\centering
\includegraphics[width=0.9\textwidth]{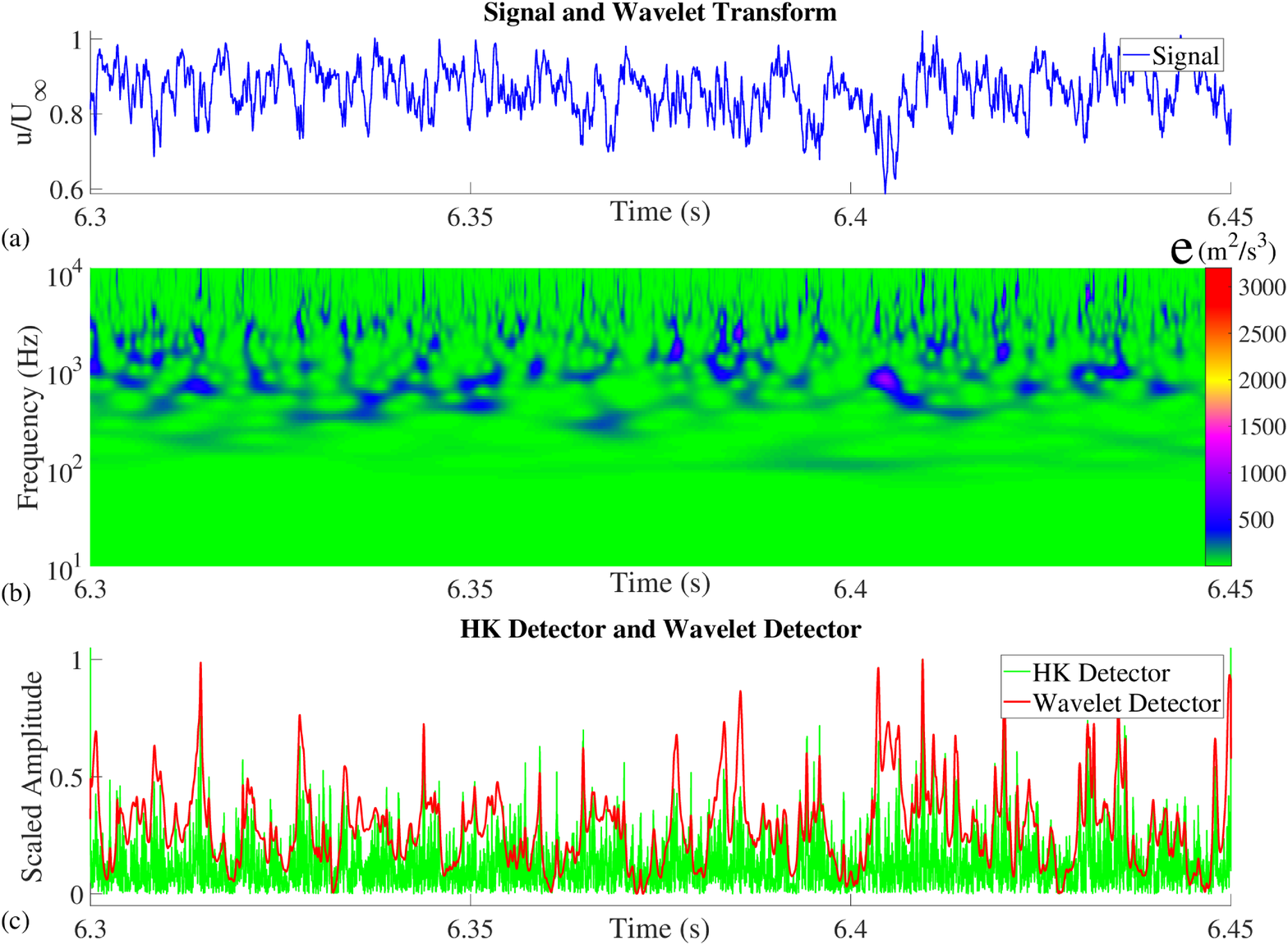}
\caption{Canonical Turbulent Boundary Layer; $y/\delta$=0.49. (a) Velocity signal $(u/U_\infty)$ vs. time (s). (b) Contour plot of (pre-multiplied) wavelet energy ($fC_w^2)$ in frequency-time plane. (c) Comparison between the wavelet detector \textcolor{red}{(red line)} and the Hedley-Keffer (HK) detector \textcolor{green}{(green line)}. Both the detectors are scaled from 0-1 over the time interval shown.}\label{Canonical_p49}
\end{figure*}

\begin{figure*}[h]%
\centering
\includegraphics[width=0.9\textwidth]{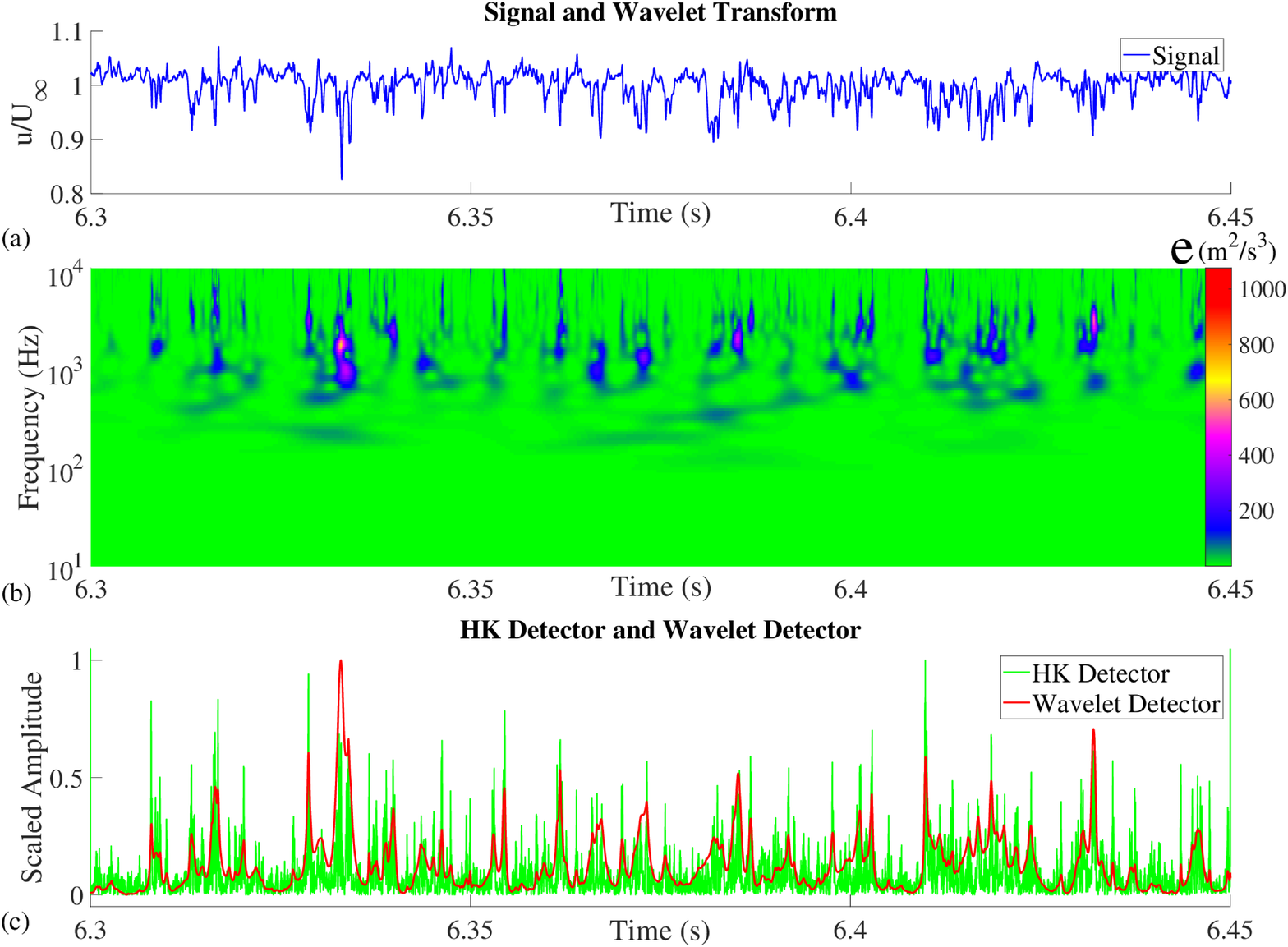}
\caption{Canonical Turbulent Boundary Layer; $y/\delta$=0.9. (a) Velocity signal $(u/U_\infty)$ vs. time (s). (b) Contour plot of (pre-multiplied) wavelet energy ($fC_w^2)$ in frequency-time plane. (c) Comparison between the wavelet detector \textcolor{red}{(red line)} and the Hedley-Keffer (HK) detector \textcolor{green}{(green line)}. Both the detectors are scaled from 0-1 over the time interval shown. The effects of edge intermittency are apparent in the signal.}\label{Canonical_p9}
\end{figure*}

Figure \ref{Compare_Indicator} shows a comparison between the indicator functions obtained from the HK criterion and wavelet detector functions using a common threshold as discussed above. There is a reasonably good match between the two indicator functions; the small deviations are expected due to the different types of detector function involved. This is also evident in the variation of the intermittency factor plotted in Fig. \ref{Indi} for different stages of transition. There is a good match between $\gamma$ values calculated based on the wavelet detector and HK criterion functions. The maximum difference is about 2$\%$ which is small compared to the typical uncertainty involved in accurate determination of intermittency factors \citep{canepa2002experiences}.

Note that in the above we have used the Hedley-Keffer method (along with visual inspection) for determining the indicator function for illustrative purposes. The wavelet detector function proposed herein should work with any other method used for setting the threshold and determining the indicator function.

\section{Wavelet detector for intermittent signals in canonical wall turbulence}

To test the generality of the wavelet detector function, we apply it to a totally different scenario involving intermittent velocity signals representing the edge intermittency in a canonical turbulent boundary layer (TBL). The experimental setup used for this purpose is different from that used above for measuring transitional velocity signals (although the wind tunnel facility is the same). The TBL was generated on a flat plate with a sharp leading edge and velocity profile was measured at a distance of 1 m downstream of the leading edge, where conditions corresponding to canonical wall turbulence  were established. Measurements were carried out at $U_{\infty}=17.5$ m/s corresponding to a Reynolds number of $Re_\theta = U_{\infty}\theta/\nu=2600$, where $\theta$ is the momentum thickness. More details on the experimental setup and the mean and root-mean-square profiles for the TBL can be found in \cite{Abhishek2022LEBD}.

For a fully turbulent boundary layer, the intermittency factor is expected to be unity (Table \ref{tab1}) over the bulk of the flow. However, as the edge of the boundary layer is approached the effects of edge intermittency start becoming apparent \citep{Kovasznay1970}, which is reflected as intermittent laminar patches in an otherwise turbulent velocity signal. Therefore the detector function designed to pick out turbulent spots in a transitional boundary layer should also be able to pick out laminar patches associated with the edge intermittency. (Incidentally, \cite{hedley_keffer_pdf_intermittency1974turbulent} originally demonstrated their method on the intermittent velocity signals near the edge of a TBL.) To test this expectation we have chosen velocity signals at two heights within the canonical TBL measured by \cite{Abhishek2022LEBD}: $y/\delta=0.49$ and $y/\delta=0.9$. The first location corresponds to a fully turbulent signal and the second corresponds to an intermittent signal; see Figs. \ref{Canonical_p49}(a) and \ref{Canonical_p9}(a). The distribution of (pre-multiplied) wavelet energy for $y/\delta=0.49$ (Fig. \ref{Canonical_p49}(b)) is similar to that for the ``turbulent'' case discussed in section 4 (Fig. \ref{Turbulent}). On  the other hand, the wavelet energy for $y/\delta=0.9$ shows extended regions of low activity corresponding to the presence of laminar patches in the signal; Fig. \ref{Canonical_p9}(b). This behaviour is qualitatively similar to that for the transitional velocity signals seen in Figs. \ref{Transition_Onset}-\ref{Nearly_Turbulent}. The only difference is that the frequency range over which the wavelet energy is intermittent is shifted to higher frequencies for the TBL, as the TBL measurement is performed at a higher freestream velocity ($U_{\infty}=17.5$ m/s) than that used for the transitional boundary layer ($U_{\infty}=7.5$ m/s). The wavelet and Hedley-Kefer detector functions for the two signals are compared in Figs. \ref{Canonical_p49}(c) and \ref{Canonical_p9}(c). For both the signals, the wavelet detector is seen to smooth out the fluctuations in the HK detector and provides an approximate envelope for the latter. The wavelet detector shows large values corresponding to the turbulent regions and extended regions of low magnitude when the laminar patches are present (Fig. \ref{Canonical_p9}c). These are desirable qualities of a detector function which the proposed detector adequately satisfies.

\section{Conclusion}\label{sec4}

We have proposed a wavelet-transform based detector function as a substitute for the derivative-based criterion function commonly used for calculating intermittency for transitional velocity signals. The subjectivity involved in specifying smoothing period in determining criterion function is brought out using the absolute value of second derivative of velocity as the detector function (consistent with the Hedley-Keffer method) and averaging it over different smoothing periods. Continuous wavelet transform of the velocity signals is carried out and a new detector function is obtained by averaging the pre-multiplied wavelet energy over all frequencies. This choice of the detector is motivated by the observation that the non-turbulent parts of an intermittent velocity signal are represented as extended regions of low wavelet energy in the frequency-time plane. The wavelet detector shows much less fluctuations than the Hedley-Keffer detector and acts as an approximate envelope over the latter. This makes the wavelet detector sufficiently smooth thereby minimizing fake dropouts and obviating the need to specify a smoothing period. At the same time the detector provides a good contrast between the turbulent and non-turbulent regions, enabling easier discrimination between the two. The distinguishing features of the proposed detector are as follows.

\begin{itemize}
    \item The averaging is performed along the frequency axis and not the time axis.
    \item The specification of averaging interval is not arbitrary as averaging is done over the \textit{entire} frequency range.
\end{itemize}

The effectiveness of the wavelet detector is demonstrated by  calculating intermittency factors for transitional velocity signals measured in our previous experiment on roughness-induced transition. The indicator functions obtained from our detector  compares well with those obtained from the Hedley-Keffer criterion function (with $T_s=100\Delta t$) based on common threshold levels; the resulting intermittency factors also match well between the two methods. Next, we have used the wavelet detector to characterize the intermittent velocity signals observed in canonical TBL measured in a different experimental setup. The wavelet detector picks out the laminar patches near the edge of the TBL fairly well suggesting that the proposed detector should work in any general situation.

The wavelet detector function can, in principle, be used with any method that prescribes threshold for obtaining an indicator function. We believe this feature will prove useful in engineering situations (e.g. wind- and gas-turbine blades) for an accurate calculation of intermittency factors.

\section*{Declarations}

\section*{Ethical approval}
Not applicable.

\section*{Competing interests}
The authors declare no competing interests.

\section*{Authors' contributions}
SSD conceived of the study. SD and AA carried out the data processing and analysis. SD prepared the figures. SD, AA and SSD  wrote the manuscript. SSD supervised the work.

\section*{Funding}
We acknowledge financial support from Science and Engineering Research Board (SERB), India (Grant No.: ECR/2018/002417) towards conducting experiments on the turbulent boundary layer.

\section*{Availability of data and materials}
Data are available from the corresponding author (SSD) on request.

\bibliography{sn-bibliography}

\end{document}